\documentclass[aps,prl,twocolumn,groupedaddress]{revtex4-1}
\usepackage[pdftex]{graphicx}
\usepackage{amsmath}
\usepackage{amssymb}

\begin{document}
\setlength\topmargin{-0.5in}
\setlength\oddsidemargin{-0.37in}

\title{Reversible Fluorination of Graphene: towards a Two-Dimensional Wide Bandgap Semiconductor}

\author{S.-H. Cheng}
\affiliation{Department of Physics, The Pennsylvania State
University, University Park, PA 16802}
\author{K. Zou}
\affiliation{Department of Physics, The Pennsylvania State
University, University Park, PA 16802}
\author{F. Okino}
\affiliation{Department of Chemistry, Faculty of Textile Science and Technology, Shinshu University, 3-15-1 Tokido, Ueda 386-8567, Japan}
\author{H. R. Gutierrez}
\affiliation{Department of Physics, The Pennsylvania State University, University Park, PA 16802}
\author{A. Gupta}
\affiliation{Department of Physics, The Pennsylvania State University, University Park, PA 16802}
\author{N. Shen}
\affiliation{Department of Physics, The Pennsylvania State University, University Park, PA 16802}
\author{P. C. Eklund}
\affiliation{Department of Physics, The Pennsylvania State University, University Park, PA 16802}
\author{J. O. Sofo}
\affiliation{Department of Physics, The Pennsylvania State University, University Park, PA 16802}
\author{J. Zhu}
\affiliation{Department of Physics, The Pennsylvania State University, University Park, PA 16802}


\begin{abstract}
We report the synthesis and evidence of graphene fluoride, a two-dimensional wide bandgap semiconductor derived from graphene. Graphene fluoride exhibits hexagonal crystalline order and strongly insulating behavior with resistance exceeding 10 G$\Omega$ at room temperature. Electron transport in graphene fluoride is well described by variable-range hopping in two dimensions due to the presence of localized states in the band gap. Graphene obtained through the reduction of graphene fluoride is highly conductive, exhibiting a resistivity of less than 100 k$\Omega$ at room temperature. Our approach provides a new path to reversibly engineer the band structure and conductivity of graphene for electronic and optical applications.
\end{abstract}

 \maketitle
\section{Introduction}
Graphene is widely considered a promising material for future electronics. The ability to tailor its properties, especially the opening of a gap in its band structure, is critical to fulfill this potential. Conventional material growth techniques, such as doping and alloying are either difficult to implement or incompatible with the desire to preserve its high electrical conductivity. Current efforts towards gapping graphene follow mainly two routes, i.e. quantum confinement in nanoribbons~\cite{Han2007,Li2008,Chen2007} and chemical modification of the graphene plane through oxidation~\cite{Stankovich2006,Stankovich2007,Gijie2007,GomezNavarro2007,Ruoff2008,Wu2008,Kaiser2009} and hydrogenation~\cite{Sofo2007,Ryu2008,Elias2009}. While considerable progress had been made using each approach, challenges remain before either can produce high-quality materials suitable for electronics applications. Even graphene nanoribbons made using the state of the art lithographic tools or chemical routes show variable width along the same ribbon~\cite{Han2007,Li2008,Chen2007}. Graphene oxide (GO) produced by the widely used Hummer's methods possesses excellent exfoliation, solubility and materials application potentials\cite{Stankovich2006,Stankovich2007,Gijie2007,GomezNavarro2007,Ruoff2008,Wu2008,Kaiser2009} but is inherently amorphous~\cite{Mkhoyan2009} due to multiple oxygen bonding configurations. In contrast, a complete fluorination of graphene can produce a two-dimensional crystal graphene fluoride, which is predicted to have a band gap of $\sim$ 3.5 eV\cite{Charlier1993}. Along a similar route, a complete hydrogenation of graphene will produce graphane, another carbon-based wide bandgap material\cite{Sofo2007}. Wide bandgap semiconductors such as GaN and SiC are widely used in high-power electronics and light emitting devices\cite{Morkoc1994}.  Graphene fluoride and graphane, if realized, will be the thinnest among this group and may be more readily integrated with carbon electronics. Although encouraging first steps have been taken, the synthesis of graphane remains challenging\cite{Ryu2008,Elias2009}. In the literature, fluorine chemistry has been used as an effective tool to modify the chemical and structural properties of carbon materials such as graphite, nanotubes and fibers, which led to a range of applications such as nanocomposites, gas storage and catalysis\cite{Touhara2000,Khabashesku2002,Lee2007}. The conductivity of fluorinated carbon is much less studied. Pristine graphene fluoride has not been produced\cite{Worsley2007}. 

In this work, we synthesize multi-layer graphene fluoride and examine its structural and electronic properties. Graphene fluoride possesses hexagonal crystalline order with an in-plane lattice that is roughly 4.5\% larger than that of graphene and exhibits strongly insulating behavior with a room temperature resistance in excess of 10 G$\Omega$, consistent with a large band gap. Multi-layer graphene regenerated through the reduction of graphene fluoride exhibits resistivity of less than 100 k$\Omega$ at room temperature. Our experiments provide a new method to reversibly control the band structure of graphene.
\section{Synthesis and structural characterizations}
\begin{figure*}
\includegraphics[angle=0,width=6in]{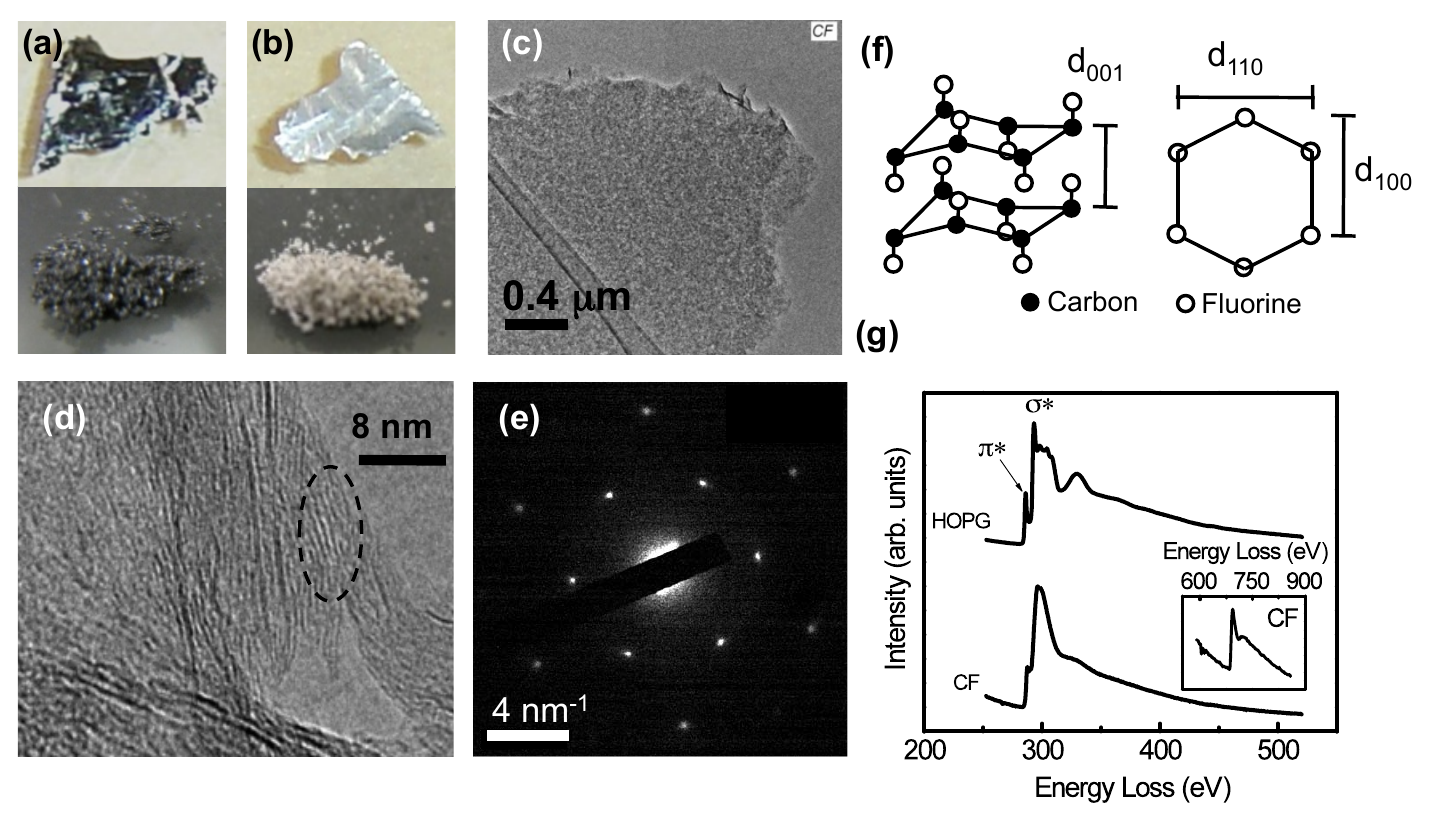}
\caption[]{(Color online) Optical images of HOPG (top) and SP-1 graphite (bottom) before (a) and after (b) fluorination. (c) Bright-field TEM image of a thin HOPG fluoride sheet suspended on a grid. (d) Cross-sectional view of a folded corner of a suspended HOPG fluoride sheet. The circled area corresponds to an interlayer spacing of 6.4 \AA. (e) Electron diffraction pattern of a suspended HOPG fluoride sheet. (f) Schematics of graphite fluoride. (g) EELS data showing the K-edge of carbon atoms in thin sheets of HOPG and fluoride. Inset: The K-edge of fluorine atoms in fluoride. The energy dispersion was 0.2 eV/channel, and the absolute energy loss scale was calibrated using the graphitic $\pi ^{*}$ peak at 285 eV. 
\label{sample}}
\vspace{-0.2in}
\end{figure*}

 \begin{figure}
\includegraphics[angle=0,width=3.0in]{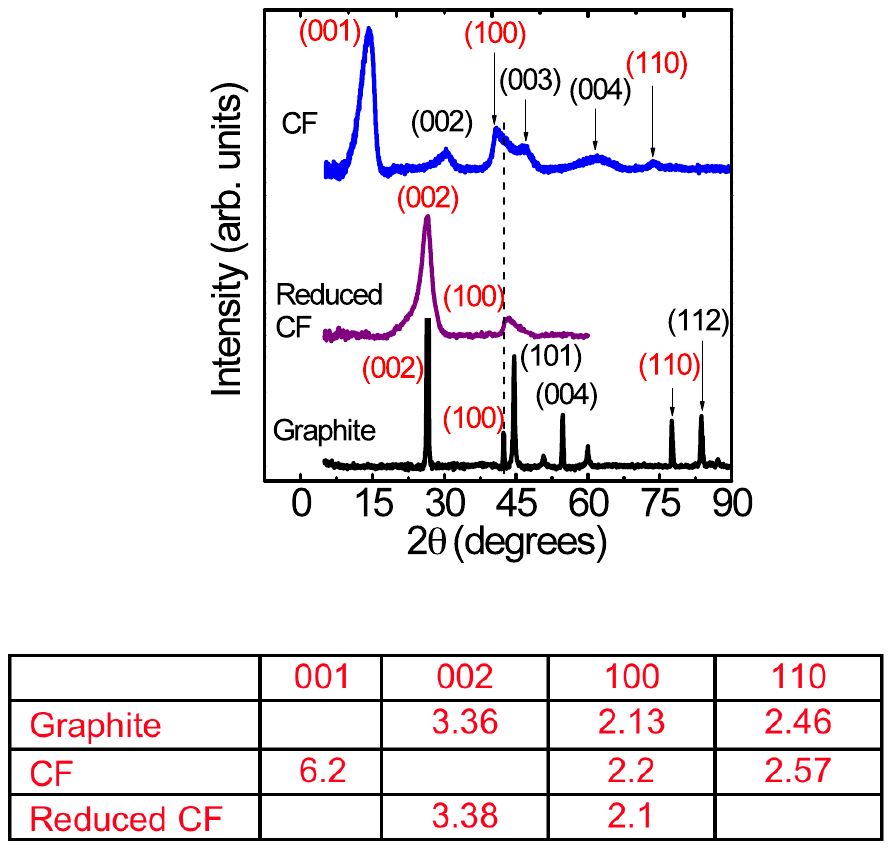}
\caption[]{(Color online) X-Ray diffraction data of fluorinated SP-1 graphite, reduced SP-1 graphite fluoride, and SP-1 graphite. The Miller indices of the diffraction peaks are marked in the figure. The dashed line is a guide to the eye.  
\label{XRD}}
\vspace{-0.2in}
\end{figure}

We follow established methods to synthesize graphite fluoride by reacting Highly Ordered Pyrolytic Graphite (HOPG) (SPI supplies, ZYA grade) or SP-1 graphite (Union Carbide, purity 99.4\%, average particle diameter 10 $\mu$ m) with fluorine\cite{Touhara2000}. Fluorine gas was introduced at 115$^{\circ} $C initially. The reaction was run for ca. 36 - 48 hours at 600 $^{\circ}$C in 1 atm fluorine. Fig.~\ref{sample}(a) and (b) show the optical images of HOPG and SP-1 graphite before and after the fluorination. The white color of the fluorinated compound provides the initial evidence of high-degree fluorination\cite{Touhara2000}. We estimate the ratio of F/C to be roughly 0.7 for the fluorinated HOPG by weight gain measurements and elemental analyses indicate the ratio of F/C to be close to 1 for the fluorinated SP-1 graphite. 
Similar to the structure of graphite, graphite fluoride consists of layers of graphene fluoride (CF). Each carbon atom in graphene fluoride is in a sp$^{3}$ bonding configuration. The position of the covalently bonded fluorine atom alternates between "up" and "down", resulting in the buckling of the in-plane C-C bonds\cite{Watanabe1988}. A shematic of the crystal structure of graphite fluoride is given in Fig.~\ref{sample}(f). To characterize the structure of our fluorinated compounds, we exfoliate the fluorinated HOPG crystals into thin sheets in the solution of isopropanol using ultrasound sonication and subsequently immobilize the sheets onto a Cu grid with a lacey-carbon support. Multi-layer graphene samples are made with identical methods and used as reference samples. Transmission Electron Microscope (TEM) measurements are made in a JEOL-2010F microscope equiped with a Gatan Enfina TM 1000 Electron Energy Loss System (EELS). 

\begin{table}
\caption{\label{tab:table1}
Lattice spacings (in unit of \AA) in SP-1 graphite, fluorinated SP-1 graphite and reduced SP-1 fluoride corresponding to the diffraction peaks in XRD data.}
\begin{ruledtabular}
\begin{tabular}{|c|c|c|c|c|}
 Index & (001) & (002) & (100) & (110) \\
\hline
Graphite & & 3.36 & 2.13 & 2.46 \\
\hline
CF & 6.2 & & 2.2 & 2.57 \\
\hline
Reduced CF & & 3.38 & 2.1 & \\
\end{tabular}
\end{ruledtabular}
\end{table}

Fig.~\ref{sample}(c) shows a bright-field TEM image of a thin HOPG fluoride sheet suspended on a grid. In contrast to the smooth surface and straight edges of graphene, CF sheets show irregular edges and spotted contrast. This contrast may be due to thickness, density or composition variations. This appearance is consistent with the partial fluorination of the HOPG fluoride sample shown in the top half of Fig.~\ref{sample}(b), where domains of CF$_{x}$ with different x may coexist\cite{Enoki2003}. Fig.~\ref{sample}(g) plots the EELS spectra of the K-edge of carbon atoms in a suspended HOPG fluoride sheet, together with that of a graphene sheet. The $\pi ^{*}$ peak corresponds to the excitation of the 1s electron to the empty $\pi ^{*}$ orbitals. This peak appears prominently in graphene (top trace) but is strongly suppressed in graphene fluoride (bottom trace), indicating the lack of $\pi ^{*}$orbitals. This data supports the formation of covalent C-F bonds in our HOPG fluoride samples. In Fig.~\ref{sample}(d), we show the cross-sectional view of a folded corner of a suspended HOPG fluoride sheet. Parallel lines are due to (001) planes of the HOPG fluoride sheet. The characteristic interlayer spacing d$_{001}$ is approximately 6.4 \AA (circled area), in good agreement with x-ray diffraction (XRD) data shown in Fig.~\ref{XRD}. Variations occur, which may be due to the partial fluorination and the possible existence of intercalated fluorines in this sample\cite{Enoki2003}.

\begin{figure*}
\includegraphics[angle=0,width=6in]{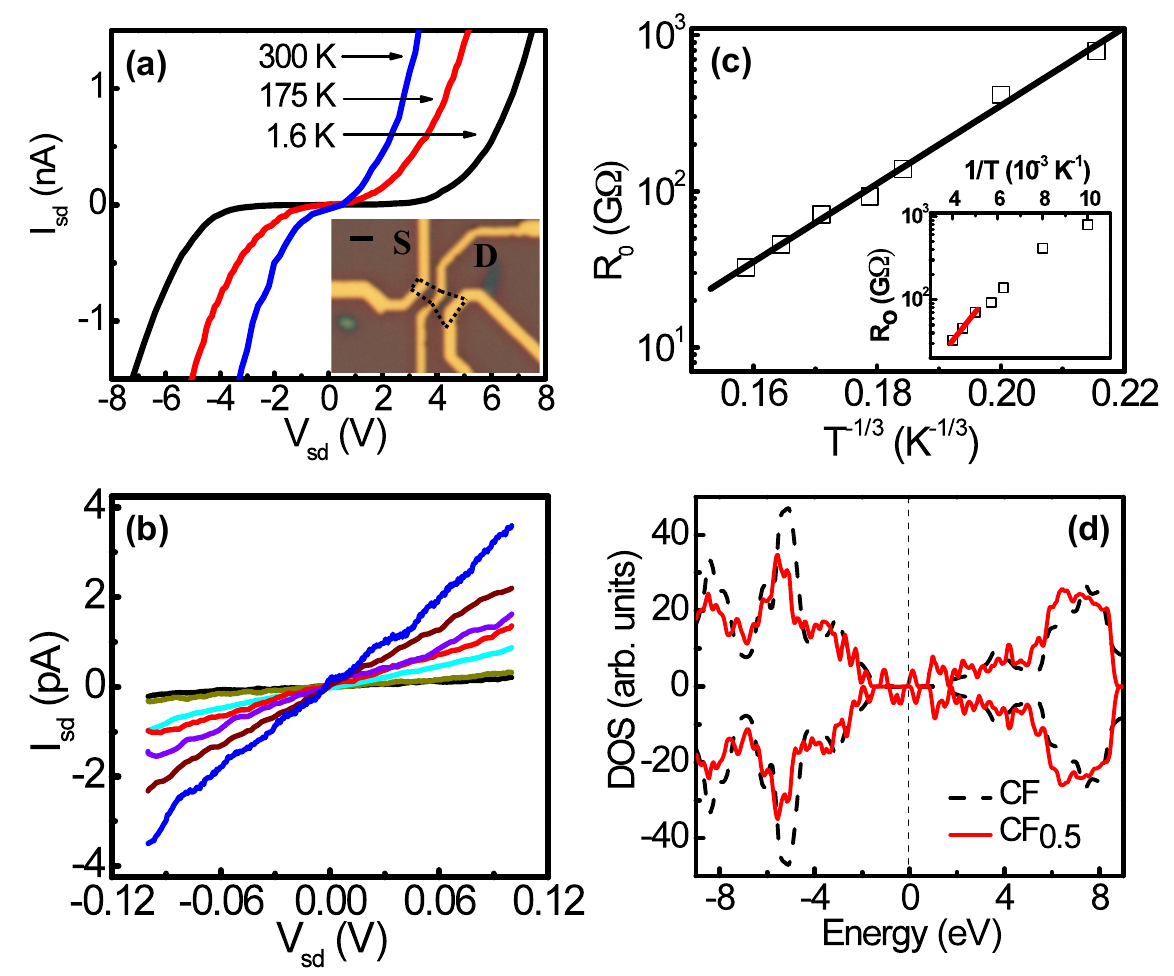}
\vspace{-0.1in}
 \caption[]{(color online) (a) I-V characteristics of a multi-layer graphene fluoride sample at selected temperatures as indicated in the figure. Inset: the optical image of the sample with the source and drain electrode marked. The graphene fluoride sheet is outlined in dotted lines. The channel is approximately 1 $\mu$ m from source to drain (center to center). Scale bar is 1 $\mu$ m. (b) I-V curves of the same sample near $V_{\mathrm{sd}}$=0V. From top to bottom: T = 250 K, 225 K, 200K, 175 K, 160 K, 125 K, 100 K. (c) Semi-log plot of the zero-bias differential resistance $R_{\mathrm{0}}$ vs. $T^{-1/3}$ of the sample in (a). Inset: Semi-log plot of $R_{\mathrm{0}}$ vs. $T^{-1}$. (d) The DoS for a random configuration of fluorine atoms attached to graphene at a 50\% concentration. Spin down is shown as negative values. The mid-gap states correspond to the p$_{z}$ orbitals from the carbon atoms not attached to fluorine. The DoS of CF (dashed line) is shown as a reference. The Fermi level is used as zero energy.
\label{CF}}
\vspace{-0.2in}
\end{figure*}
In Fig.~\ref{sample} (e), we show the electron diffraction (ED) pattern of a suspended HOPG fluoride sheet. Both the first and second order diffraction spots are clearly visible, indicating in-plane hexagonal crystalline order. From the first order diffraction spots, we find the in-plane lattice distance d$_{100}$ (Fig.~\ref{sample}(f)) of this sheet to be 2.4\% larger than that of the reference graphene sheets. Such expansion is expected as C-C bonds in graphene convert from sp$^2$ to sp$^3$ configurations in CF\cite{Watanabe1988}. The magnitude of the expansion is smaller than determined from XRD data shown in Fig.~\ref{XRD}. This is also reasonable considering the partial fluorination of this sample.

XRD patterns of the nearly 100\%~fluorinated SP-1 graphite powder (bottom of Fig.~\ref{sample}(b)) are given in Fig.~\ref{XRD}, together with that of the SP-1 graphite and the reduced SP-1 graphite fluoride powder. The extracted lattice spacings are given in Table~\ref{tab:table1}. For fluorinated graphite, the interplanar distance of d$_{001}$=6.2 \AA~agrees well with the TEM image shown in Fig.~\ref{sample}(d). Its in-plane lattice constant d$_{110} $ is 4.5\% larger than that of graphite. These results are in very good agreement with the literature \cite{Sato2004} and provide strong evidence to the crystalline nature of graphene and graphite fluoride.

\section{Transport measurements}
Previous density functional theory (DFT) calculations show that graphite fluoride possesses a gap of approximately 3.5 eV at the $\Gamma$ point of its band structure. The bands are weakly dispersed in the z direction due to very weak interlayer coupling~\cite{Charlier1993}. Our band structure calculations of monolayer graphene fluoride confirm this large gap, which may further increase when quasi-particle corrections are included. A large band gap in CF may enable digitial transistor and optical applications currently unavailable in prinstine graphene. This aspect of CF is hardly explored in the literature. To test its electrical properties, we obtain multi-layer graphene fluoride sheets (6-10 nm) by mechanically exfoliating fluorinated HOPG to SiO$_{2}$ substrates and fabricate devices using conventional e-beam lithography followed by metal depositions. The metal electrodes consist of 5nm Cr and 40 nm Au. An optical image of a two-terminal CF device is given in the inset of Fig.~\ref{CF}(a). Transport measurements are carried out in a pump He$^{4}$ cryostat with a variable temperature range of 1.4-300 K. Two-terminal conductance is obtained by applying a constant dc bias V$_{\mathrm{sd}}$ (Keithley 2400 or Yokogawa 7651) across the sample and measuring the resulting current I$_{\mathrm{sd}}$ using a current preamp (DL1211). Fig.~\ref{CF}(a) shows the $I_{\mathrm{sd}}$($V_{\mathrm{sd}}$) (I-V) of a multi-layer CF device at selected temperatures in the range of 1.6 K $< T <$ 300 K. $I_{\mathrm{sd}}$($V_{\mathrm{sd}}$) is highly non-linear at large $V_{\mathrm{sd}}$ and decreases rapidly with decreasing temperature. These I-V curves are reproducible in forward and backward V$_{\mathrm{sd}}$ sweeps without hysteresis. Fig.~\ref{CF}(b) plots $I_{\mathrm{sd}}$($V_{\mathrm{sd}})$ in the range of -0.1 V $< V_{\mathrm{sd}} <$  0.1 V at different temperatures. $I_{\mathrm{sd}}$($V_{\mathrm{sd}}$) is approximate linear in this regime, the slope of which yields the zero-bias differential resistance $R_{\mathrm{0}}$ =d$V_{\mathrm{sd}}$/d$I_{\mathrm{sd}}$. $R_{\mathrm{0}}$($T$) measures approximately 30 G$\Omega$ at 250 K and increases with decreasing temperature, reaching 790 G$\Omega$ at 100 K. The same measurement setup without the sample displays an ohmic leakage resistance of approximately 1.8 T$\Omega$ in the whole range of V$_{\mathrm{sd}}$ shown in Fig.~\ref{CF}(a), which fluctuates slightly at different temperatures (0.2 T$\Omega$). This leakage resitance produces a parallel conduction path and has been taken account to obtain the value of $R_{\mathrm{0}}$. The large resistance of the CF device, its T-dependence and non-linear I-V all point to a strongly insulating behavior, consistent with a large band gap.

\begin{figure}
\includegraphics[angle=0,width=3in]{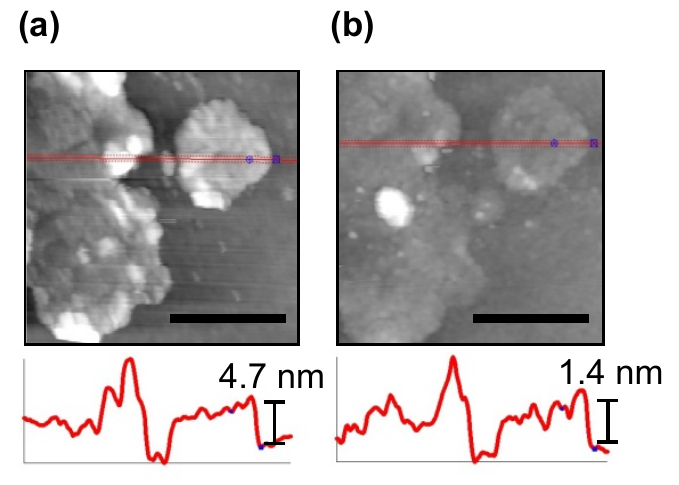}
\vspace{-0.1in}
 \caption[]{(color online) AFM images of the same sheet of fluorinated HOPG exfoliated on SiO$_{2}$ substrate before (left) and after (right) annealing in H2/Ar gas. The traces below the images correspond to the line cut across the sheet. 
\label{AFM}}
\vspace{-0.2in}
\end{figure}
In a simple band insulator, $R_{\mathrm{0}}$ follows an activated T-dependence $R_{\mathrm{0}}\propto  exp (-\Delta /2k_{B}T)$, where $\Delta$ is the band gap. To test this behavior, we plto $R_{\mathrm{0}}$($T$) vs $1/T$ in the inset of Fig.~\ref{CF}(c) in a semi-log plot. While the high-T data may be described by the exponential T-depedence, $\Delta$ = 68 meV extracted from the fit is much smaller than the expected $\Delta \sim$ 3.5 eV. Moreover, deviation from this exponential T-dependence is evident below 200 K. Interestingly, such T-dependence strongly resembles observations in disordered two-dimensional electron gas in silicon inversion layers, when the Fermi level lies in the band tail of the conduction band populated by localized impurity states. There, instead of thermal activation, carriers conduct by hopping through neighboring sites or through longer distances with the assistance of phonons (variable-range hopping)\cite{Mott1975,Tsui1974}. Such a mechanism is quite conceivable in our CF samples, where islands of graphene may remain as a result of incomplete fluorination, giving rise to localized mid-gap states. This hypothesis is supported by DFT calculations shown in Fig.~\ref{CF}(d), where we plot the density of states (DoS) of a partially fluorinated monolayer graphene fluoride CF$_{0.5}$. In this simple scenario, 50\% of the fluorine atoms are randomly missing. As a result, the p$_{z}$ orbitals of sp$^2$ bonded carbon atoms produce finite a DoS inside the band gap of CF. These states can participlate in the conduction via hopping mechanisms. We expect this scenario to be applicable in partially fluorined multi-layer graphene fluoride as well.

\begin{figure}
\includegraphics[angle=0,width=3in]{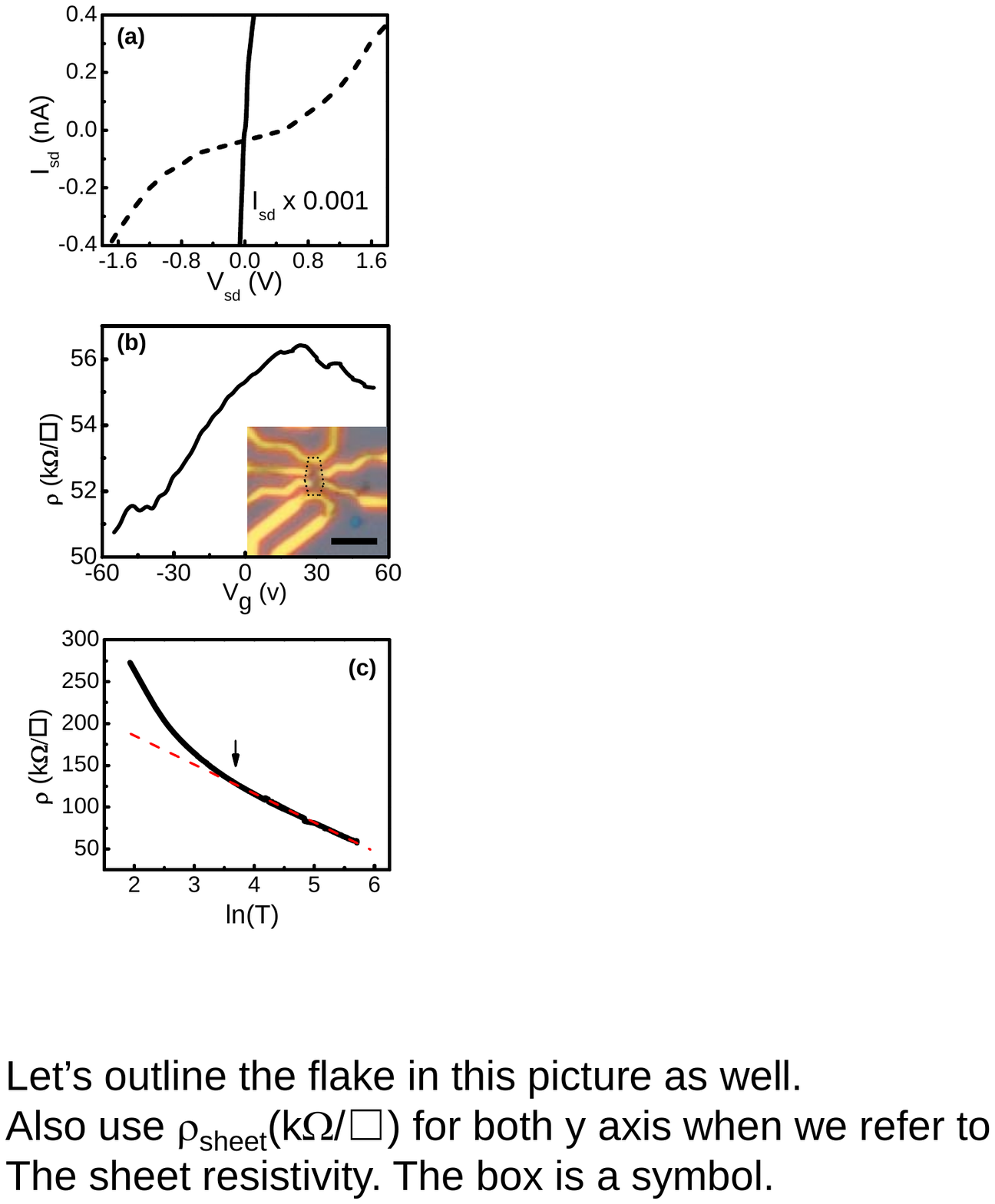}
\vspace{-0.1in}
 \caption[]{(color online) (a) I-V characteristics of reduced graphene fluoride (solid) in contrast to CF (dashed). $I_{\mathrm{sd}}$ of reduced CF has been reduced by a factor of 1000 to fit the scale. (b) Sheet resistance vs backgate voltage V$_{\mathrm{g}}$ of a reduced CF sample showing a charge neutrality point at V$_{\mathrm{g}}$ = 24 V with a maximum resistance of 56 k$\Omega$. T=130 K. Inset: typical multi-terminal reduced CF device on SiO$_{2}$/doped Si substrate. The edge of the piece is outlined in dotted line. The scale bar is 4 $\mu$ m. (c) The sheet resistance $\rho$ ($T$) vs. $ln$($T$) of a reduced CF sample in the temperature range of 5 $< T <$ 300 K. The red dashed line is a guide to the eye. The arrow indicates 40 K. $V_{\mathrm{g}}$ = 0 V.
\label{graphene}}
\vspace{-0.2in}
\end{figure}
To test the validity of the variale-range hopping (VRH) model, we plot $R_{\mathrm{0}}$ vs. $T^{-1/3}$ of the CF sample in  Fig.~\ref{CF} (c). The linear trend shows that $R_{\mathrm{0}}$ ($T$) is well described by the $exp [(T_{0}/T)^{1/3}]$ functional form corresponding to VRH conduction in two dimensions\cite{Mott1975}. Although other exponents cannot be ruled out by the data itself, we note that due to their thickness (6-10 nm) our samples are well into the 2D regime even at room temperature so a 2D exponent is reasonable. We extract $T_{\mathrm{0}} = 1.9 \times 10^{5}$ K from the linear fit in Fig.~\ref{CF} (c). The applicability of the VRH scenario is further justified by noting that $T_{\mathrm{0}}$ is much greater than our measurement temperatures\cite{Mott1975}. The large $T_{\mathrm{0}}$ is consistent with the large resistance of CF samples and attests to the high degree of fluorination. In comparison, a similar analysis of hydrogenated graphene produces a $T_{\mathrm{0}}$ of $\sim$ 250 K\cite{Elias2009}.  

Although VRH model provides a consistent and satisfactory description of our data, given the large band gap of CF, the formation of Schottky barriers at the metal-CF interface and the contribution of contact resistance to the overall reistance of CF samples cannot be ruled out. In the literrature, Schottky barrier has been used to account for the resistance of graphene-graphene oxide juntions\cite{Wu2008}. If Schottky barriers dominated in our samples, the exponential T-dependence at high T in the inset of Fig.~\ref{CF}(c) would yield an estimated barrier height of 68 meV. This small barrier height cannot account for the large resistance observed in our samples. We therefore conclude that although it is possible that both Schottky contacts and hopping conduction coexist in our CF samples, the majority of the resistance comes from the bulk. In the literature, little is known about the alignment of CF bands with those of metals. In our samples, the position of the localized states in the band gap of CF plays an important role in determining the contact resistance as well. A clear understanding of these aspects of the material is necessary to give a comprehensive account of transport in CF but is not available at the moment. We hope that our work will inspire more future studies on this material, which will further clarify this issue. 
 
It is clear from the above measurements, however that a large gap exists in CF. It is therefore possible to use fluorination as a tool to create nonlinear, digital elements and form interconnects in a graphene circuit. Combined with patterning, graphene nanochannels surrounded by insulating CF may be created as an alternative to nanoribbons.  Adding to the flexibility of this approach, we now demonstrate the reversibility of the fluorination reaction by regenerating graphene from graphene fluoride. In the literature, CF is shown to be stable in vacuum up to 500 $^{\circ}$C. The reduction of CF in the presence of H$_{2}$ starts to occur at ~ 300 $^{\circ}$C\cite{Kumagai1995,Sato2006}. To remove fluorine, we anneal our CF samples in a mixture of flowing Ar/H$_{2}$ (90\%/10\%) at 500-600$^{\circ}$~C for 5 hours\cite{Kumagai1995,Sato2006}. H$_{2}$ reacts with CF to produce graphene and HF, the latter being carried away by the gas flow. Fig.~\ref{AFM} shows the AFM images of the same CF sheet exfoliated on SiO$_{2}$ before and after the annealing process. Both the height and the surface roughness of the sheet decrease significantly after annealing process, suggesting the removal of fluorine. The reduction of SP-1 graphite powder is accompanied by a color reversal from white to dark gray. XRD data show that reduced SP-1 graphite fluoride exhibits an interlayer spacing of 3.38 \AA~and an in-plane lattice spacing d$_{100}$= 2.1 \AA (Fig.~\ref{XRD} and Table 1). Both values are in very good agreement with those of graphite, attesting to the effectiveness of the defluorination process.

Multi-layer sheets (3-8 nm) are exfoliated from reduced HOPG fluoride crystals to SiO$_{2}$/doped Si substrates and made into field effect transistors using lithographic techniques described earlier. The I-V curve of a reduced CF sample exhibits ohmic behavior with a resistance of $\sim$ 130 k$\Omega$ at 300 K, which is approximately 5 orders of magnitude smaller than that of the CF device shown in Fig.~\ref{CF}(a), plotted here as a dashed line for comparison. Four-terminal measurements are carried out on devices such as shown in the inset of Fig.~\ref{graphene}(b) using standard low-frequency lock-in techniques with an excitation current of 50-200 nA. Fig.~\ref{graphene}(b) shows the sheet resistance $\rho$ vs. the backgate voltage V$_{\mathrm{g}}$ of a reduced CF sample, where a maximum of 56 k$\Omega/\square$ is reached at V$_{\mathrm{g}}$=24 V. The sheet resistance is generally below 100 k$\Omega$ at 300 K and is smaller than the resistance of graphene reduced from GO, which are generally in the M$\Omega$ range\cite{Gijie2007,GomezNavarro2007}. We attribute the reasonably high conductivity of reduced CF to the crystalline nature of graphene fluoride. Reduced CF, however, is still $\sim$100 times more resistive than pristine graphene exfoliated from bulk graphite. The broad XRD peaks of CF and reduced CF shown in Fig.~\ref{XRD} suggest that defects are created in the fluorination/reduction processes, possibly due to the high temperatures involved (600 $^{\circ}$C). This may have been the major reason behind the loss of conductivity in reduced CF. We also cannot rule out the possibility that the reduction process may not have been complete. 

The $T$-dependent sheet resistance $\rho$($T$) of a reduced CF device is given in Fig.~\ref{graphene} (c). $\rho$($T$) increases from 60 k$\Omega$ at 300 K to 270 k$\Omega$ at 7 K. Unlike reduced GO\cite{GomezNavarro2007,Kaiser2009}, its T-dependence cannot be described by the VRH model. Empirically, it follows an approximate $ln T$ dependence for 40 $< T <$ 300 K and becomes more strongly T-dependent at lower temperature. The occurance of the transition is consist with its relatively low resistance. A detailed understanding of the conduction mechanism will be addressed in future studies. 

\section{Conclusion}
In summary, we present the synthesis, structural and electrical properties of multi-layer graphene fluoride, a two-dimensional wide bandgap semiconducting material derived from graphene. We demonstrate that fluorination and reduction reactions can reversibly modify the conductivity of graphene by many orders of magnitude. Our initial results show that crystalline graphene fluoride can be synthesized. It has a large band gap and exhibits strongly insulating transport while graphene regenerated through its reduction exhibits reasonably high conductivity. This method complements the existing chemical routes towards band structure engineering of graphene. With suitiable fluorine-resistant and/or heat-resistant substrates and electrodes, spatially selective fluorination and reduction may be carried out on graphene and graphene fluoride devices, offering the opportunity to craft functional elements in a graphene circuit. As a wide bandgap material, graphene fluoride may lead to electro optical applications. These opportunties, as well as the improvement of the crystal quality, will be explored in future experiments. 

\begin{acknowledgments}
We are grateful for helpful discussions with Tom Mallouk and Greg Barber and technical assistance from Mark Angelone. This work is supported by NSF MRSEC DMR-0820404, NSF CAREER DMR-0748604 and the American Chemical Society Petroleum Research Fund. The authors acknowledge use of facilities at the PSU site of NSF NNIN and the Penn State Materials Simulation Center.
\end{acknowledgments}

%

\end{document}